\documentstyle[aps,multicol,epsf]{revtex}
\input{epsf}

\begin{document}

\title{Correlation length by measuring empty space in simulated aggregates}
\author{R. M. L. Evans$^*$ and M. D. Haw}
\address{Dept.~of Physics and Astronomy, University of Edinburgh, EH9
3JZ, U.K. \\
$*$Current address: Dept.~of Physics and Astronomy, University of
Leeds, LS2 9JT, U.K.}

\date{February 12, 2002}
\maketitle

\begin{abstract}
We examine the geometry of the spaces between particles in
diffusion-limited cluster aggregation, a numerical model of 
aggregating suspensions.  Computing the distribution of distances from
each point to the nearest particle, we show that it has a scaled form
independent of the concentration $\phi$, for both two- (2D) and
three-dimensional (3D) model gels at low $\phi$. The mean remoteness
is proportional to the density-density correlation length of the gel,
$\xi$, allowing a more precise measurement of $\xi$ than by other
methods.  A simple analytical form for the scaled remoteness
distribution is developed, highlighting the geometrical information
content of the data. We show that the second moment of the
distribution gives a useful estimate of the permeability of porous
media.

 \vspace{-9mm}
\end{abstract}



\begin{multicols}{2}

Substantial attention has been paid to cluster-cluster aggregation as
a computational model of gelation in colloidal suspensions, aerogels,
aggregating emulsions, etc.~\cite{ourrev}.  There is a large
literature analysing the structure of individual aggregates, in terms
of fractal geometry \cite{fracDLCA,Vicsek}.  However it is now
properly understood \cite{ourrev,Gimelsolgel} that, in irreversibly
aggregating systems at non-zero particle concentration $\phi$,
clusters, while fractal at intermediate length scales, must inevitably
become space-filling at the largest scales, as clusters pack to form a
`gel' with a well defined correlation length $\xi$.  Direct analysis
of the gel structure remains rare \cite{GimelPRB}, although it is
fundamental to a gel's macroscopic physical properties.  Colloidal
gels and aerogels are porous media and, as such, the geometry of their
`pores' or vacancies is of prime importance, influencing the flow of
permeating solvents and gel stability under external stresses such as
gravity \cite{Laura,bookflock}.  Here, using the diffusion-limited
cluster aggregation (DLCA) model \cite{fracDLCA}, we examine the
structure of the space inside the porous gel.  DLCA is perhaps the
simplest realistic model of aggregation in colloidal suspensions, and
has been shown to qualitatively reproduce many features of
experimental systems \cite{ourrev,Gonzalez95,Bibette92}.  We show that
examination of the void structure in the simulated gel leads to a
measure of the correlation length $\xi$ of unprecedented precision,
while the scaling with $\phi$ also gives an estimate of the cluster
fractal dimension at scales smaller than $\xi$. We further find that,
at low $\phi$, the pore structure is independent of $\phi$, i.e. DLCA
particle gels have statistically identical structure regardless of
particle concentration.

In fractal aggregates, there is no single well-defined pore size,
since spaces of all sizes are present. To investigate the {\em
distribution} of void sizes we require a suitable measure; but it is
difficult to avoid some arbitrariness in delineating the extent of a
void \cite{Allain91,Schwarzer94}.  Marine cartographers face a similar
problem: where does one sea end and the next begin? We bypass the
problem as follows. If an adventurer adrift in a boat sights land
nearby, (s)he cannot be sure whether (s)he is on a small lake, or near
the edge of a large ocean. However, if (s)he knows that the nearest
land is far away, (s)he can infer the existence of a large body of
water.  Hence information on the distribution of void sizes can be
gained by measuring the {\em remoteness}, i.e.~the distance to the
nearest particle, of all points in an aggregated system. This
distribution of remoteness should {\em not} be interpreted, as in
Ref.~\cite{Ehrburger-Dolle91}, as the distribution of void sizes.  The
former is only an indirect measure of the latter.

The DLCA algorithm is implemented as follows. A square (2D) or cubic
(3D) grid, with periodic boundary conditions, is occupied by a
randomized population of particles at concentration $\phi$. Each
particle performs a random walk on the lattice until it finds a
particle on a neighbouring site. The two then stick irreversibly.  A
cluster of particles thus formed continues its random walk, at a rate
inversely proportional to its radius of gyration, forming ever larger
aggregates on contact with neighbouring clusters. The growing flocs
have a fractal geometry characterised by a dimension $d_f$ which is
lower than the space dimension $d$, so that, on the scale of a cluster
of `mass' $M$ and radius $r$, the mean concentration ($\sim M/r^d \sim
r^{d_f-d}$) decreases as the cluster grows. When the clusters grow to
such a size ($\xi$) that their mean concentration ($\sim\xi^{d_f-d}$)
matches the overall system concentration $\phi$, mass conservation
demands that the clusters must fill space. Assuming a
concentration-independent fractal dimension (evidenced in the limit of
low $\phi$ \cite{Lach-hab96,Bibette92}), the correlation length in the
`gels' varies as $\xi\propto\phi^{-1/(d-d_f)}$. To avoid artifacts in
the simulations, the correlation length $\xi$ must be both
significantly larger than the lattice parameter $a$, and smaller
\cite{Ehrburger-Dolle91} than the system size $L$, $L\gg\xi\gg a$.

We analyse the `gel' structure after all particles have
aggregated. The fact that the particle positions are confined to the
lattice is irrelevant when finding the distribution of remoteness; we
visit all points in the system, and measure the {\em Euclidean}
distance to the nearest particle (contrast Ref.~\cite{Schwarzer94}). 
For each gel, we compile a histogram of remoteness $h(R)$, which is
normalized so that $h(R)\,\mbox{d}R$ is the probability that a point
chosen at random has a remoteness in the range $R\pm\mbox{d}R/2$.  It
is significant that we find $h(R)$ to be normalizable. This is not the
case for
\end{multicols}

\twocolumn
\noindent
similar analyses of isolated fractal clusters
\cite{Allain91,Schwarzer94,Dubuc89}, where an artificial boundary
must be imposed as the empty space surrounding a cluster is, in
principle, limitless. The distributions we measure contain a natural
cut-off, revealing that no points in the system are more remote than
some finite distance.

\begin{figure}
\begin{center}
  \epsfysize=8cm\leavevmode\epsffile{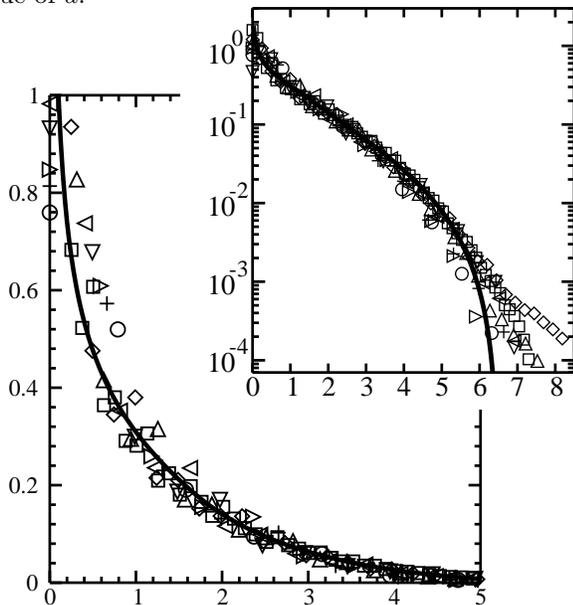}
	\caption{For simulations in $d=2$, the distributions,
	$p(x)$, of remoteness in units of the mean remoteness, for
	concentrations $\phi=0.10\,\Box$, $0.15\,\Diamond$,
	$0.17\,\triangle$, $0.20\,\triangleleft$,
	$0.23\,\bigtriangledown$, $0.25\,\triangleright$, $0.27\,+$,
	$0.30\,\bigcirc$. Solid line: one-parameter theoretical fit.
	{\bf Inset:} The same graph on a log scale and over a
	wider range, making the tail of the distribution more visible.}
\label{2D}
\end{center}
\end{figure}

\begin{figure}
\begin{center}
  \epsfysize=8cm\leavevmode\epsffile{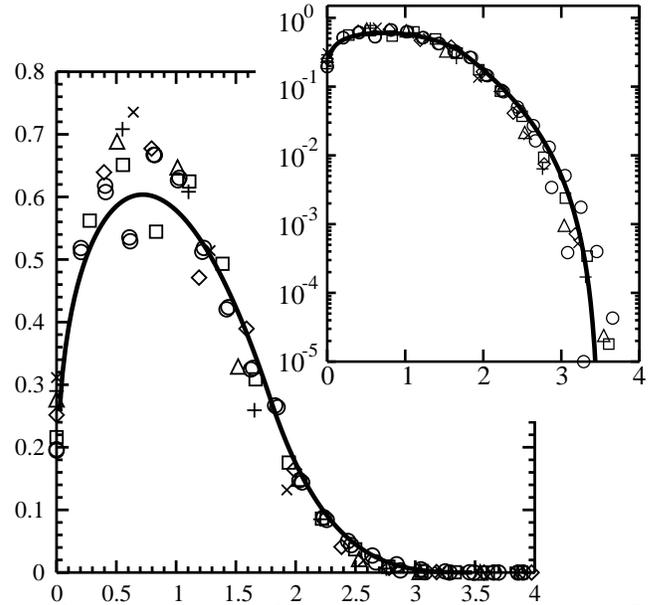}
	\caption{As Fig.~\protect\ref{2D}, for $d=3$, at
	concentrations $\phi=0.02\,\bigcirc$, $0.03\,\Box$,
	$0.05\,\Diamond$, $0.07\,\triangle$, $0.08\,+$,
	$0.10\,\times$.}
\label{3D}
\end{center}
\end{figure}

Given the hypothesis that the correlation length is the only relevant
length scale present, the cut-off distance must be a fixed fraction of
$\xi$, as must other lengths such as the mean remoteness $\langle
R\rangle$.  To check whether a single length scale characterises the
geometry at various $\phi$, we re-scale each histogram by its mean
$\langle R\rangle$, and define a scaled remoteness distribution $p(x)$
with unit mean and unit norm, by
\mbox{$p(x)=\langle R\rangle\,h(\langle R\rangle x)$}, where $x$ is
remoteness measured in units of the mean (Figs.~\ref{2D} and \ref{3D}
for $d=2$ and $d=3$ respectively). Points at the smallest values of
$x$ are affected by grid artifacts, but otherwise, the data collapse
very well onto a single master curve for each value of $d$.
\vspace{-4mm}
\vspace{-4mm}

\begin{figure}
\begin{center}
  \epsfxsize=6cm\leavevmode\epsffile{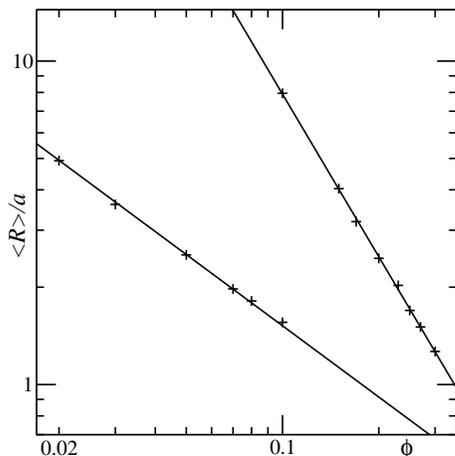}
	\caption{Log-log plot of $\langle R\rangle/a$ vs.~$\phi$ 
	for 2D (upper set) and 3D (lower set) simulations.}
\label{meanR}
\end{center}
\end{figure}

{\em No fitting} was required to achieve the data collapse in
Figs.~\ref{2D} and \ref{3D}, since the scale factor $\langle R\rangle$
is a well defined property of the gel. Furthermore, when we plot this
length scale, the mean remoteness, against concentration $\phi$ in
Fig.~\ref{meanR}, we obtain remarkably clear power-law behaviour.  As
discussed, a power-law relationship between $\phi$ and any length
scale proportional to $\xi$ is expected at these low concentrations,
but the almost complete absence of noise was quite unforeseen. Compare
equivalent plots in the literature \cite{Gonzalez95} using the length
scale obtained from the peak in the structure factor. We deduce that
moments of the remoteness distribution are precise, noise-free
measures of the length scale present in the gel.

Further confirmation that $\langle R\rangle$ is a measure of, though
much smaller than, the correlation length $\xi$ comes from the
gradients of the lines in Fig.~\ref{meanR}, which are respectively
$-1.67\pm0.05$ and $-0.73\pm0.03$ for $d=2$ and $d=3$. Given our expected 
gradient $-1/(d-d_f)$, this implies a fractal dimension
$d_f=1.40\pm0.02$ for $d=2$, in agreement with the accepted
value of 1.42 (see e.g.~\cite{ourrev}), and $d_f=1.63\pm0.05$ for $d=3$,
marginally below the accepted zero-concentration value of 1.80.
\vspace{-2mm}
\vspace{-3mm}

To find the constant of proportionality relating $\langle R\rangle$ to
the more commonly used measure of the correlation length, from the
peak in the structure factor at wavevector $q_{\rm max}$, we
compute the ratio 
$\langle R\rangle/\xi=\langle R\rangle q_{\rm max}/2\pi$ for each
gel, and find a constant value $0.034\pm0.004$ for $d=2$, and
$0.121\pm0.007$ for $d=3$. We also measure higher moments of the
remoteness distribution (table \ref{moments}).

To understand the shapes of the master curves in Figs.~\ref{2D} and
\ref{3D}, we take as our starting point the fact that the gelled
aggregate is self-similar on scales smaller than the correlation
length.  On some scale, we may picture the structure as a set of
amorphous blobs and cavities.  These cavities contain points with a
range of remoteness, characterised by a normalized remoteness
distribution $f(R)$. The precise form of $f(R)$ depends on the shape
of the cavity. If we `{\em zoom in}' to a smaller length scale
$\sigma$, we notice that the blobs are themselves aggregates of blobs
and cavities. By self-similarity, those smaller cavities have a
normalized remoteness distribution $f(R/\sigma)/\sigma$, i.e.~of the
same shape as the larger cavities, but scaled down. Thus, compiling
the overall histogram of remoteness, $h(R)$, we get contributions
proportional to $f(R/\sigma)$ from all scales $\sigma$, up to a
largest scale $\chi\sim\xi$. The magnitude of the contribution from
each length scale is as yet undetermined. Let us call it $c(\sigma)$.
Hence,
\begin{equation}
\label{h}
	h(R) = \int_0^\chi c(\sigma)\, f(R/\sigma)\, \mbox{d}\sigma.
\end{equation}
We have expressed $h(R)$, the remoteness distribution for a complex
fractal aggregate, in terms of $f(R)$, the distribution of remoteness
in some `primal' cavity with a simple (albeit unknown), non-fractal
geometry. Equation (\ref{h}) may be regarded as a definition of
$f(R)$.

Since the geometry is self-similar on all length scales $\sigma$ in
the integration, $0\leq\sigma\leq\chi$ (ignoring a lower cut-off due
to the grid), $c(\sigma)$ must be a
function with no characteristic length scale, i.e.~a power of $\sigma$
up to the cut-off $\chi$. In fact, we have no prior justification for
the cut-off to be sharp since, in principle, scale-invariance could be
lost gradually over a range of scales up to $\chi$. 
Nevertheless, the hypothesis will
be validated by the data.  We define
the as yet undetermined exponent $\nu$ by
\mbox{$c(\sigma)=c_0\,\sigma^{\nu-2}$}. The constant coefficient $c_0$
is fixed by the normalizations of $h(R)$ and $f(R)$ as
$c_0=\nu\chi^{-\nu}$. As mentioned, the distribution of
remoteness $h(R)$ is not synonymous with the distribution of void
sizes. We may identify $c(\sigma)$ as the latter, while $f(R)$
contains information on the void {\em shapes}.

Recasting (\ref{h}) in terms of the scaled master curve
$p(x)=\langle R\rangle\,h(\langle R\rangle x)$, and 
putting $z=\langle R\rangle x/\sigma$, yields
\begin{equation}
\label{p}
	p(x) = \frac{\nu}{x} \left(\frac{x}{x_0}\right)^\nu
		\int_{x/x_0}^\infty z^{-\nu} \, f(z) \, \mbox{d}z
\end{equation}
where $x_0\equiv \chi/\langle R\rangle$, which is not a free parameter:
from Eq.~(\ref{h}),
$x_0^{-1} = \langle x\rangle_f\,\nu/(\nu+1)$
where $\langle x\rangle_f$ is the mean of the distribution $f(x)$.
Without loss of generality, we set the size scale of the cavity
described by $f(x)$ so that 
\linebreak
\vspace{-5mm}
\vspace{-5mm}
\begin{table}[h]
	\begin{tabular}{|c|cccc|}
	$d$	&
		$\left.\langle R^2\rangle^{1/2}\right/\langle R\rangle$ &
		$\left.\langle R^3\rangle^{1/3}\right/\langle R\rangle$ &
		$\left.\langle R^4\rangle^{1/4}\right/\langle R\rangle$ &
		$\left.\langle R^5\rangle^{1/5}\right/\langle R\rangle$ \\
	\hline
	2 & 1.42(3) & 1.77(5) & 2.07(8) & 2.35(10) \\
	3 & 1.141(9) & 1.25(2) & 1.35(3) & 1.44(3)
	\end{tabular}
	\caption{Ratios of the moments of the remoteness distribution,
	averaged over all simulations (error in last digit).} 
\label{moments}
\end{table}

\noindent
the most remote point within it is separated by unity from the nearest
boundary. Hence $f(x)=0$ for $x>1$, implying $p(x)=0$ for $x>x_0$. So
we interpret $x_0$ as the point on the $x$-axis of Figs.~\ref{2D} and
\ref{3D} at which the graph falls to zero, i.e.~the value, in units of
$\langle R\rangle$, of the natural upper cut-off in the remoteness
distribution.

The value of the parameter $\nu$ can be calculated by examining the
small-$x$ behaviour of Eq.~(\ref{p}).  An object closely related to
the remoteness distribution is the Minkowsky cover
\cite{Schwarzer94,Dubuc89,bookfrac}, the locus of points lying within
a fixed distance, $R$ say, of a fractal set.  For small $R$, the
volume of the Minkowsky cover grows as $R^{d-d_f}$
\cite{Schwarzer94,Dubuc89,bookfrac}.  Our distribution $h(R)$ is
proportional to the perimeter of the Minkowsky cover, i.e. the
derivative of its volume with respect to $R$. Hence, we require
$p(x)\propto x^{d-d_f-1}$ in the limit of small $x$. As $x\to0$, the
integral in Eq.~(\ref{p}), assuming it is convergent, tends to a
constant, so that $p(x)\propto x^{\nu-1}$ and we identify $\nu=d-d_f$.

It remains only to determine the form of the primal distribution
$f(x)$, i.e.~the geometry of the basic cavity. We know it is normalized
and must vanish for
$x>1$. We shall discuss some further features but, because
Eq.~(\ref{p}) already embodies many of the correct properties for a
gelled aggregate, we find the results to be insensitive to the precise
form of the function $f$.

Let us examine how $f(x)$ vanishes as $x$ approaches unity from
below. Consider a circular cavity with unit radius in 2D, inside which
are drawn contours of equal distance from the edge,  i.e.
concentric circles. The cavity's remoteness distribution $f(R)$ is
proportional to the length of the contour at a distance $R$ from the
edge. Hence, as we approach the centre of the cavity, $R\to1$,
$f(R)$ falls {\em linearly} to zero. This would also be the case, in
2D, for a triangular cavity, or any regular polygon. In the centre
(most remote part) of a 3D cavity, on the other hand, the 
distribution vanishes quadratically, since it is proportional to the
area of an equal-remoteness surface. Hence, one might consider
substituting the normalized distribution $f_{\rm trial}=(1-x)^{d-1}d$ for
$0<x<1$ into Eq.~(\ref{p}), thus building the fractal texture from
a set of regular `lagoons' of all sizes. 
This formula was {\em not} used for the curves in Figs.~\ref{2D} and
\ref{3D}, despite good agreement with the 2D data, as
it has shortcomings in 3D. For $d=3$, $\nu\approx 1.2$
so that, as $x\to0$, the integral in Eq.~(\ref{p}) diverges if $f(0)$
is finite. At small $x$, we demand $p(x)\sim
x^{d-d_f-1}$, which requires $p$ to be an {\em increasing} function of
$x$ for $d=3$, where $d_f\approx1.8$.  When we
construct a surface of constant distance around a tree-like `branch',
near the branch the area of the surface must {\em grow} as the
distance from the branch is increased, because the branch is {\em
convex}. But the negative gradient of $f_{\rm
trial}$ describes only {\em concave} lagoons. According to our data,
such a function builds an adequate picture of
the spaces in a 2D aggregate, since the aggregate with $d_f\ge1$ is 
sufficient to partition 2D space
into concave lagoons of all sizes.  But in 3D,
branches form an imperfect {\em cage} which has significant
 {\em convex} structure.

To acknowledge the existence of both concave and convex parts of the
aggregate, we use the simplest possible function $f(x)$ that vanishes
in the correct way (with a power $d-1$) both at $x=0$ and $x=1$.  We
construct a piecewise smooth function that is the lower envelope of
two polynomials: $c_1 x^{d-1}$ and $c_2 (1-x)^{d-1}$ with constant
coefficients $c_1=\beta^{1-d}d$ and $c_2=(1-\beta)^{1-d}d$ chosen to
ensure that it is continuous and normalized. The position of the cusp
where the two parts meet, at $x=\beta$, is the only free parameter of
our model. The mean of this function is at 
$\langle x\rangle_f=(1+[d-1]\beta)/(d+1)$ so that we may eliminate 
the parameter $\beta$ in favour of $x_0$ which we fit to the measured
value of the cutoff in the master curves. 

We find best values $x_0=6.5\pm1.0$ and $3.5\pm0.2$ for $d=2$
and $3$ respectively.  These one-parameter fits are the curves shown
in Figs.~\ref{2D} and \ref{3D}. Our single formula reproduces the data
very accurately both for $d=2$ and $d=3$, even in the tails of the
distributions, until the data become
noisy.

The second moment of $h(R)$ has a special significance to porous
media. At low Reynolds' number, the mean flow rate, $\langle
u\rangle$, of a solvent of viscosity $\eta$ through a porous medium
subject to a pressure drop per unit length $\nabla P$, is given by
Darcy's law \cite{Darcy}, $\langle u \rangle = k \nabla P / \eta$,
where the permeability $k$ is a squared length scale that depends only
on the geometry of the porous medium. To avoid complex hydrodynamic
calculations, $k$ is often estimated by the Carman-Kozeny (CK) relation
\cite{Schwartz93}, $k\approx(1-\phi)^3/5S^2$, in terms of the specific
surface area $S$. However, since the flow velocity at any given point
is not as dependent on the total surface present as it is
on the distance to the nearest obstacle, the CK relation
mis-calculates the porosity in materials with more than one
characteristic pore size. For instance, the flow rate through a set
of parallel cylindrical holes, of various radii $a$, can be calculated
exactly using Poiseuille's formula \cite{Poisseuille}, yielding
$k=(1-\phi)\langle a^4\rangle/8\langle a^2\rangle$, which,
expressed in terms of the remoteness, is $k=3\langle
R^2\rangle/4$. By substituting the surface area of the holes into the
CK formula, one obtains $k_{\rm CK}\approx(1-\phi)\langle
a^2\rangle/10$, which approximates the exact value only for a sharply
peaked distribution of radii, for which $\langle
a^4\rangle\approx\langle a^2\rangle^2$. The approximation fails too
for a polydisperse set of parallel-sided channels, where
the exact result can be expressed as
$k=\langle R^2\rangle/2$. We therefore propose an approximate rule of
thumb $k\approx 0.6\langle R^2\rangle$, which continues to hold for
polydisperse pores and for rough surfaces while, in both cases, the
CK equation fails. The new formula has been shown to yield
results of the correct order of magnitude for fractal particulate gels
\cite{Laura}. However, like the CK equation,
the new formula lacks information on pore topology, so must
also fail in certain cases, e.g.~where some pores are closed. There is
then no alternative to a full hydrodynamic computation.

The shape of the remoteness distribution carries information on the
geometry in the DLCA simulations, and confirms (by scaling onto a
master curve) that, statistically speaking, all the simulations at the
low concentrations considered, give rise to the same geometry: the
particle gel `pore' structure is independent of concentration. 
The mean remoteness, an easy quantity to measure in the
simulations, is proportional to the correlation length (at least in
the range of concentrations studied), hence giving a measure of the
gel's characteristic length scale that is considerably less noisy, and
perhaps more easily determined, than other measures. The {\em second} moment
of the remoteness distribution provides an
estimate of the permeability of the gel. These coordinate-independent
measures of real-space length scales could be applied to
a range of other phenomena, such as the void geometry in {\em
reaction}-limited aggregation \cite{Warren91} and in earlier
stages of aggregation \cite{Gimelsolgel}, or the $\phi$-dependence
at higher concentrations \cite{Lach-hab96} when the structure is no
longer controlled solely by the zero-concentration critical phenomena.

RMLE acknowledges the support of the Royal Society and the Royal
Society of Edinburgh.

\end{document}